\documentclass{ws-mpla}
\begin{document}

\markboth{} {}

%
\catchline{}{}{}{}{}
%

\title{Dark matter, dark energy and modern cosmology: the case for a
Kuhnian paradigm shift}

\author{\footnotesize J. E. Horvath$^*$}

\address{Instituto de Astronomia, Geof\'\i sica e Ciencias Atmosf\'ericas\\ Rua do
Mat\~ao 1226, 05508-900 S\~ao Paulo SP, Brazil\\
$^*$foton@astro.iag.usp.br\\}

\maketitle

\pub{Received (Day Month Year)}{Revised (Day Month Year)}

\begin{abstract}

Several works in the last few years devoted to measure fundamental
probes of contemporary cosmology have suggested the existence of a
delocalized dominant component (the "dark energy"), in addition to
the several-decade-old evidence for "dark matter" other than
ordinary baryons, both assuming the description of gravity to be
correct. Either we are faced to accept the ignorance of at least
95 \% of the content of the universe or consider a deep change of
the conceptual framework to understand the data. Thus, the
situation seems to be completely favorable for a Kuhnian paradigm
shift in either particle physics or cosmology. We attempt to offer
here a brief discussion of these issues from this particular
perspective, arguing that the situation qualifies as a textbook
Kuhnian anomaly, and offer a tentative identification of some of
the actual elements typically associated with the paradigm shift
process "in the works" in contemporary science.

\end{abstract}

\section{Introduction}

Thomas S. Kuhn (1922-1996) in the 20th century imprinted a strong
pattern under which scientific research is seen today. Even
philosophers, historians and epistemologists which disagree with
his views about these subjects still find difficult to avoid a
discussion for or against Kuhn's own framework (see, for example,
S. Fuller  ' Is There Philosophical Life after Kuhn?´ , Philosophy
of Science, v. 68, 2001, 565-572$^{1}$).

In his book$^{2}$ The Structure of Scientific Revolutions  the
author discussed in a long essay style the basic concepts and
operating mechanisms of the scientific enterprise, quite often
resorting to a normative viewpoint. Scientific progress is seen
mainly as a succession of paradigm shifts between periods of
"normal science", inside which the task of the scientists is
rather to confirm and reinforce the existing paradigms. The
boundaries of these "normal science" periods have been termed by
him scientific revolutions, truly extraordinary episodes in the
research history, triggered by the repeated failure in solving a
(big) problem(s) in the field and/or a new discovery shaking the
very field foundations and not easily fitted into the existing
paradigm. The latter concept may be in turn defined the sum of the
theories and value commitments shared by the scientific group,
later rephrased provisionally as "discipline matrix" for this
specific meaning. According to this definition, the scientific
groups are bound by theories but also other elements (concepts,
procedures and even symbolic generalizations usually called "laws"
such as Newton's ${\overrightarrow{f}} = m \times
{\overrightarrow{a}}$ and a similar entities), constituting the
common grounds on which research is conducted. Scientific research
is thus seen from a common context (gestalt), and it is only when
the efforts to fit a problem/phenomenon into the paradigm fail
repeatedly that "extraordinary science" sets in, and is accepted
(or rather, tolerated) by traditionalists in search of a more
satisfactory understanding. A lot of criticism has been published
against these ideas, and sometimes bold extrapolations of them
constructed for application in other fields, like public policies
and pedagogy. In addition, the Kuhnian perspective has been
recognized as akin to Darwinian evolution, or rather to the stasis
theory of Eldredge and Gould$^{3}$ postulating punctuated
equilibrium of biological evolution instead of a gradual and
continuous change of life forms.

Cases which may be considered textbook examples of the paradigm
shift are known in several sciences (although never without some
dispute). They range from truly big, ground-shaking revolutions
such as the well-known Copernican and Newtonian; to smaller and
more specialized events like the emergence of gauge principles in
field theory. A more recent possible example, to which this work
is devoted, is the case of modern cosmology in which one set of
new facts is being widely discussed and seeking for a
comprehensive global picture, still absent or very blurred.
Because of its importance we shall outline the scientific case in
some extent (but keeping technical details to a minimum) in the
next section, with emphasis to the connections to previous ideas
and results. The accelerated Universe

Quite recently the interest in astrophysics and cosmology bloomed
boosted by the advance of technological facilities, and allowed a
series of studies which reached and captured the imagination of
the public opinion. Specifically, cosmology has been highlighted
by the reports from 1998 on about the acceleration of the universe
seen in studies of type Ia supernovae $^{4,5,6}$ with an
indication of a non-zero value of a delocalized component known as
"dark energy" (hereafter dark energy) as a possible (but not
unique) solution.

The argument for suck a remarkable claim is as follows. Type Ia
supernovae form a class of stellar explosions long associated to
the death of an "old" evolved star. This general statements relies
on the fact that, in contrast to other explosive events (known as
type Ib, or type II) hydrogen is absent in the ejected gas.
Therefore, it is concluded that the exploding star had exhausted
the hydrogen and hence, it must have evolved from the
hydrogen-burning phase well before the event. What is a crucial
step, and forms the basis of the cosmological analysis is the
contention that type Ia supernovae are quite homogeneous as long
as their absolute brightness is considered, and therefore form a
set of standard candles. In addition, a remarkable relation
between the maximum brightness and a time interval defined
properly from the rise of the lightcurve to its decline has been
discovered$^{7}$, a feature that allows a further calibration of
the astronomical magnitudes (that is, to infer the absolute
brightness and to put a distance for each source).

When these explosions are observed in distant galaxies, affected
by the expansion of the universe as discovered by Hubble and
confirmed in several detailed works, their distances inferred by
looking at the lightcurves can be compared to the distance to the
same galaxy as inferred by the observations of the position of the
atomic element lines (which gives the so-called redshift, long
attributed to the very expansion of the substratum of spacetime).
The claims by the dedicated groups can be rephrased as the
assertion that distant supernovae are systematically fainter than
they "should be" according to the constant Hubble flow. Hence,
either supernovae were intrinsically less brilliant in the past,
or the expansion has accelerated, and that is why they look dimmer
than expected. Reasons to support the first possibility could be
(and have been) advanced, but they were dismissed on observational
grounds (for example, some "dust" component absorbing light should
do so in the same amounts for each wavelength, something
completely at odds with all types of actual cosmic dust
observations). The data independently gathered by two different
teams do show the same behavior and thus constitutes a
cross-checked evidence for the proposed accelerated expansion.

Strong as this evidence seems, it is still reinforced by a similar
feature independently inferred from the combined data of the WMAP
experiment and other initiatives measuring background radiation
maps$^{8,9}$. The experiments actually measure tiny temperature
fluctuations in the cosmic fluid which decoupled from the rest of
matter at the time in which the photons ceased to scatter off
charged particles, at the era of hydrogen recombination.
Recombination physics is quite well-known and the fraction of
ionized hydrogen can be calculated with confidence, and even
estimated from first principles. This happens quite early in the
primitive expanding universe, at a time around 300 000 yr after
the "Big Bang" itself, giving rise to an almost-perfect black body
radiation (in fact, by far the best measured in Physics) if not
for these tiny irregularities mentioned above. However, these are
precisely the inhomogeneities (of the temperature and therefore of
the matter density field coupled to it at the very early universe)
which are believed to grow well after the hydrogen recombination,
to eventually form galaxies and the structure of the present
universe. It is by measuring the pattern of these fluctuations
that the contribution of each component to the total energy
density of the universe can be gauged. Generally speaking,
cosmologists refer to these components in term of fractions of the
closure density, a numerical value that would make the universe to
have exactly zero curvature (or, more loosely, an exact balance of
all components to produce a simple geometry). The measurements of
the cosmic background radiation are strikingly compatible with the
sum amounting to this critical value $\Omega_{tot} = 1$ (where
$\rho_{i}/\rho_{c} \equiv \Omega_{i}$ is the referred fractional
contribution of the i-th component to the closure density, and
thus $\Omega_{tot} = \sum \Omega_{i}$). However, the direct
counting of visible components do not amount to much more than
$\Omega_{ordinary \, matter} \sim 0.04$ , a result also limited
from above by the element abundances of primordial
nucleosynthesis$^{10}$. Adding the dark matter component (see
below), the total matter content of the universe must be
$\Omega_{total \, matter} \sim 0.25$ . Yet, the difference between
the total matter content and the cosmic background radiation
inference calls for a dominating component, precisely in the
amount needed to explain the supernova data as well (as long as it
can do the job of producing the required acceleration, which
additionally requires quite a special relationship between its
energy density and pressure).

These recent reports pointed to a problem that should be added to
the ancient "dark matter" one, namely, the existence of a
clustered component mostly of non-baryonic origin which adds
another substantial fraction of the matter-energy content balance.
Actually, this proposal dates back to the decade of 1930, when
astronomer F. Zwicky compared the matter directly "seen" in the
form of stars and gas residing in the galaxy$^{15}$ with the
matter needed to hold the system together. Since the former fell
short by a factor ~ 5-10 to do the job, he concluded that most of
the matter was not producing light, and was then "dark" (in fact
Zwicky firmly believed that these "dark particles" must be
ordinary like protons and nuclei, therefore he rather spoke of
"missing light"). Later, similar arguments based on observations
were elaborated by many researchers, till the point that the "dark
matter" problem became part of the disciplinary matrix of
astronomy, rejected by a few and unsolved for several decades. We
shall return to this point below to review how astronomers reacted
to this situation and its possible relation to the newer "dark
energy" fact.

Several possible alternatives for both dark matter and dark energy
unexplained components are being considered by the
cosmologists/particle physicists communities, the solutions
ranging from "conservative" to "wild" approaches. No full
solution, and in fact not even a firm hint of it is still
available. In this situation we may legitimately wonder whether we
are witnessing a scientific revolution "in the works", or if the
problems could be rather solved within the existing concepts and
theories. We attempt to offer here a brief discussion of these
issues, with a tentative identification of some of the actual
elements typically associated with the paradigm shift process.

\section{Standard cosmology: Friedmann - Robertson - Walker
models}

After a somewhat lengthy development in the first half of the 20th
century, the discovery of the cosmic microwave background
radiation$^{11}$ , primordial nucleosynthesis$^{10}$  and
large-scale structure$^{12}$ studies helped to shape what is
called today the "standard" cosmology. The resonant success of
General Relativity as a theory of gravitation prompted its
application to the largest self-gravitating system of all, the
Universe itself. For that purpose, the available data suggested,
and a sensible theoretical thinking indicated, the adoption of the
so-called Cosmological Principle. This statement is generally
expressed as follows: the Universe looks the same in all
directions and has no privileged position.

In conjunction with the General Relativity framework, the
Cosmological Principle serves to select a set of homogeneous and
isotropic solutions (known as Friedmann-Robertson-Walker
cosmologies) in which the dynamics is described by the Friedmann
equations$^{13}$. The latter equations relate the scale factor of
the universe $a(t)$ to the content of matter, radiation and
whatever else composes the universe (that is, the above
$\Omega_{i}$`s), given the value of the curvature parameter
$\kappa$. Einstein equations then relate the matter-energy content
(contained in the right-hand side) to the geometric properties of
spacetime, with differential operators acting on the fundamental
object $g_{\mu \nu}$ (the metric tensor). In Wheeler's powerful
words "matter tells spacetime how to curve, and curved spacetime
tells matter how to move"$^{14}$ .

The cosmological equations of Friedmann-Robertson-Walker based on
General Relativity read

$$ {\biggl( {\dot a \over{a}}\biggr)}^{2} =
{8 \pi G \over{3}} \rho - {\kappa\over{a^{2}}} + {\Lambda\over{3}}
\,\, \, \, \, \, \, \, \, \, \, \, \; \; \; \; \;\;\;\;\;\;\;
(1a)$$

$$ {\ddot a \over{a}}= - {4 \pi G \over{3}} (\rho + 3 P)
+ {\Lambda\over{3}} \,\, \, \, \, \, \, \, \, \, \, \, \; \; \; \;
\;\;\;\;\;\;\;(1b)$$

where the "dot" indicates the derivative with respect to the
cosmic time. These equations are supplemented by a "conservation
law" which tells how the energy density $\rho$ changes with time
as the scale factor $a(t)$ evolves with time, ${\dot \rho} = -
3({\dot a}/a)(\rho + P)$. The pressure $P$ and energy density
$\rho$ of the right-hand sides are actually the sum of whatever
components contribute to them. For example, cosmic matter exerts
essentially no pressure and therefore is characterized by $P=0$ in
the above equations. Other simple cases include a radiation field
for which $P= (1/3)\rho$, and a few further known "fluids". The
important thing to retain is that the features of the components
(through $\rho$ and $P$) determine the behavior of the growing
scale factor of the Universe $a(t)$.

It was at the beginning of the 20th century that Hubble's
fundamental discovery of a linear relationship between galaxy
distance and recession velocity (later termed "Hubble law")
created significant problems for the theoretical description of
the Universe based on the triumphant General Theory of Relativity.
It is known that initially, a constant $\times$ the metric tensor
($\equiv \Lambda g_{\mu \nu}$), among the admissible terms in the
gravitational field equations, was introduced by Einstein to
produce a static universe. In fact, eqs. 1a and 1b were already
written with this contribution explicitly separated, as it can be
easily checked.

In this discussion of static vs. expanding cosmologies, it is
clear that Einstein himself seemed to dislike a non-zero $\Lambda$
possibly invoked to produce a static Universe. Eventually such a
term was deemed superfluous once the Hubble expansion was amply
confirmed, since a non-zero but very small $\Lambda$ was regarded
as a mathematically possible but physically unjustified solution.
This is a well-known documented case in the history of science.

As stated above, the recent evidence gathered on Type Ia
thermonuclear supernovae and anisotropies of the cosmic microwave
background radiation have indicated the same content of dark
energy dominating the energy balance today. Since this unclustered
form of energy should produce an accelerating phase of the
universe, the coefficient between the pressure and the energy
density must be negative (the right-hand side of eq. 1b must be
positive for${\ddot{a}} > 0$ ), something odd for normal fluids,
but not much different from a tension in a rubber band.

Given this situation, a late (contemporary) acceleration shares
some of the features postulated much earlier for a primordial
inflationary phase, and the attention has been turned to it as
well. What is inflation? Inflation is a brief, early phase of the
Universe in which the expansion rate has been much higher than any
solution based on a "reasonable" fluid dominance, in fact in many
theories the expansion of the scale factor was exponential ($a(t)
= a(t_{0}) \times exp(Ht)$ ), something that needed unusual
properties of the component dominating the Universe dynamics, not
unlike a negative relation between $P$ and $\rho$ but at a much
higher energy scale (that is, closer to the Big Bang itself).
Inflation gradually become a key ingredient in modern cosmology,
not only because it helps to solve important problems of the
observed universe (horizon, formation of structure, etc.), but
also because density perturbations generated inside it are later
greatly amplified, with a characteristic flat spectrum, and are
observed as "frozen" at the radiation-matter decoupling. In fact
the cosmic microwave background radiation data gathered today is
of high quality and permits a scrutiny of the fluctuations
generated at an early epoch in the universe. The consistency of
these analysis with inflationary predictions is very significant.
Therefore, and before turning to the issue of dark matter/dark
energy itself, we may ask first whether the evidence is strong
enough to state the Inflation itself happened.

\section{Inflationary theories: is Inflation really part of the
paradigm?}

It is perhaps significant that the very specific word "paradigm"
is now being widely used in the specialized literature to design
the latest status of the Inflation. As stated, a general
definition of the latter states that it is a (brief) period of the
universe in which the expansion is extremely fast, possibly
exponential, caused by a peculiar behavior of the equation of
state $P(\rho)$. It is certainly an elegant and neat form of
solving some serious problems related to the Big Bang cosmology,
and furthermore predicts some nice features amenable of direct
observation, such as the power spectrum of the primordial
fluctuations. Some cross-checks of these inflationary ideas,
including the power spectrum, continue to indicate the need of a
dark matter, a clustered non-baryonic component of galaxies and
galaxy clusters which has been discussed for several decades. In
fact, inflationary ideas date back to the early '80s, and are
therefore much newer than the referred  Zwicky's paper$^{15}$
pointing out the existence of dark matter in galaxies.

However, as a generic mechanism, Inflation does not offer a direct
answer to the question of the dark matter and dark energy, but
rather predicts just the total $\Omega_{tot}$ and form and shape
of $\delta \rho / \rho$, which in turn forces the existence of
some yet unknown components as a consequence of it (as mentioned
above, baryons alone are much too scarce to fill the budget). It
is very remarkable that adding up the "observed" dark matter and
the dark energy we "naturally" arrive at the inflationary
prediction value $\Omega_{tot} = 1$.

As a feature to model and understand the data, it may be stated
that Inflation is still challenged by the community, but it has
gained an ample credit lately. For many, it is now a part of the
discipline matrix. But what is important to remark is that
Inflation did not disturb the dynamics of the rest of
Friedmann-Robertson-Walker cosmology because, from a Kuhnian point
of view, it was not intended to destroy or substitute it, but
rather came to justify its initial conditions (or more precisely,
the unimportance of them) (see the related discussion in M.S.
Turner,  'Dark Matter and Dark Energy: The Critical Questions',
2002, arXiv astro-ph/0207297$^{16}$ ). Even accepting that
picture, the type of Inflation that happened and specially what
caused that Inflation (scalar fields?) are not yet answered (see
the remarks by H. Zinkernagel, 'Cosmology, particles and the unity
of science, Studies in History and Philosophy of Modern Physics,
33, 2002, 493-516$^{17}$). Therefore, while the existence of
Inflation is considered by many as part of the paradigm, its
realization rather qualifies as an unsolved problem, perhaps to be
"explained away" in the same act as the very existence of the dark
matter+dark energy if both features emerge from a still more
fundamental theory, such as braneworlds or M-Theory (see, for
example, S. Nojiri and S.D. Odintsov, 'Where new gravitational
physics comes from: M-theory?', 2003, arXiv hep-th/0307071$^{18}$)
invoking extra dimensions of the Universe. It is fair to conclude
here that the standard Friedmann-Robertson-Walker cosmology is a
much better understood framework than Inflation itself, and
despite its success the latter has been incorporated (but not yet
merged) to the former.

It is also important to remark again that Inflation is expected to
act at extremely early times only, when the universe was likely
governed by physics at the highest energies. This is a very
extreme regime, not yet probed in accelerators or laboratories,
and therefore physicists naturally entertain various ideas to
produce Inflation without actually worrying too much about the
"low-energy" Universe. In contrast, dark matter and dark energy
comprise the overwhelming majority of our everyday, steady, cold
universe, and become in this sense a matter of concern, because we
certainly should introduce them explicitly in almost every
cosmological consideration dealing with theory/data.

\section{"Invention" vs. "discovery" of $\Lambda$ and a
comparison with the history of dark matter}

As previously stated, so far a careful analysis of the
observational evidences$^{19}$ from supernovae and cosmic
microwave background radiation suggests that the "ancient"
einstenian idea of just a constant term in the field equations is
not ruled out and may be useful as a realistic model. However,
when we take a closer look, the einstenian concept of $\Lambda$ is
actually quite different from the present one. While Einstein
entertained the idea of a term $\Lambda \times g_{\mu \nu}$ as a
simple possibility allowed by symmetry criteria (and was therefore
"invented" in this sense, on theoretical grounds), we may argue
that effects of $\Lambda$ have been "discovered" in contemporary
data. It was Einstein contention to allow a term of this type on
the left-hand side (thus, attached to the geometrical content),
instead of devising some kind of fluid contributing with the same
term to the right-hand side (that is, a component of the Universe
enforcing the geometry).

The words "invention" and "discovery" are precisely the same ones
employed by Kuhn (K70) in his definition of both concepts,
exemplified by the controversy between Steele, Priestley and
Lavoisier for the priority in the discovery/understanding of
oxygen. This observation leads to question which is the actual
status of more complex models going beyond the simplest
cosmological constant, such as quintessence fields$^{20}$.
Quintessence fields are nothing but a phenomenological attempt to
introduce some dynamical component which can act as an accelerator
agent producing effectively a negative relation $P(\rho)$ as a
result of its action. The chosen name is, of course, directly
related to the aristotelic concept of the composition of the world
revived in this unexpected turn.

It is clear that we have "invented" those models, and their
obvious ad hoc character reinforces the use of this term. However,
it would not be totally out of question to speak of a "discovery",
and certainly if a particular model becomes accepted to explain
the data (say, a scalar field with some potential term), we may
hear about the "discovery" of quintessence, even if never detected
in the conventional sense. Such a hypothetical model might prove
later to be a mock manifestation of some different physical entity
(i.e. extra terms in Friedmann's equation induced by high-energy
physics). It is well-known that the recognition of a fact needs
not only data, but also its proper understanding, which in this
case is not yet achieved, and possibly lasting a finite and
unpredictable amount of time. But since the dark energy is
unlikely to be detected directly, a quite large acceptance time
may be required irrespectively of the actual outcome.

It is fair to state that, in many senses, we "see" $\Lambda$ quite
differently than Einstein did. We believe now that $\Lambda$ is
related to the zero-point energy of quantum fields, and it is
quite strange to the community that its value is orders of
magnitude smaller than the "natural" number $10^{121}$ inferred
from a simple calculation imposing the usual fluctuation behavior
of the known elementary fields. A point we would like to stress is
that the measurement of a tiny  signals a breakdown of a more
restricted paradigm ("nature manages to drive $\Lambda$ to zero"),
which reigned for several decades championed by the defenders of
the Occam's razor cosmologies. In fact, many reasons to justify
$\Lambda = 0$ were put forward prior to 1998. The small, but
non-zero value of $\Lambda$ may prove even more difficult to
justify than an exactly null figure. We do not have any reason for
such a huge mismatch between theory and observations, just as
Kepler did not have a reason for elliptic planetary orbits, later
found by Newton using his own mechanics. Perhaps a completely new
approach changes our way of looking at $\Lambda$ ,or there are
anthropic reasons to produce a tiny $\Lambda$ , but they have to
be studied and clarified$^{21}$. All this suggests again that a
new viewpoint may be needed, even if we choose to keep "standard"
gravity and succeed to identify the dark energy component.

In contrast to the case of dark energy, it is interesting to note
that the dark matter problem is almost coeval with the development
of Friedmann-Robertson-Walker  cosmologies. It is not an anomaly
appearing after that paradigmatic theory was established, but
rather a background fact, constantly reinforced and extended over
the years. However, the community eventually choose to dismiss
dark matter as a cosmological "problem" and pushed it to the realm
of Particle Physics/Astrophysics (committed to find a suitable
exotic particle/compact remnant candidate(s)). In contrast, all
issues related to dark energy have been always seen as part of the
cosmological problem. Imagine that the small but non-zero
$\Lambda$ had arisen before. Would it have anticipated the present
crisis in the standard cosmology? or it would have rather followed
the path dark matter did, namely to be considered not really a
problem, but rather an ingredient to be addressed and found by
some other related discipline?. We strongly suspect that the
second alternative would have been the one chosen, simply because
it reflects the behavior of the community when faced with an
analogous earlier situation. We believe that a small non-zero dark
energy (in its simplest "cosmological constant" incarnation)  has
now closed the room for sweeping such problems (dark matter+dark
energy) under the rug.

It is clear that, in spite of the above facts, we are not actually
claiming that there were no attempts to solve the dark matter
problem prior to the emergence of the dark energy evidence. As a
concrete example of an attempt to change the dark matter paradigm
we may cite the MOdified Newtonian Dynamics (MOND) of Milgrom et
al.$^{22}$. In this theory there is a new regime beyond a certain
acceleration scale and deviations of Newtonian dynamics happen,
for example, on galactic scales (a relativistic version that is
derivable from a Lagrangian, another key feature in the present
particle physics paradigm, has been recently presented). But now
it is clear that this kind of idea could be a solution for part of
the whole problem only, since we have to explain the existence of
the unclustered dark energy as well, and therefore they seem to be
overall less attractive than, say, a decade ago. Of course, there
is no deep crisis, just an impasse for the supporters of normal
science Inflation+Friedmann-Robertson-Walker cosmology, since for
them the dark matter+dark energy team should come as a "plug-in"
solution from the outside of their own discipline.

\section{Solving the imbroglio?}

Given the state of the art, and as a working hypothesis, we must
seriously consider the possibility the origin of the dark matter
and dark energy, and their relative contributions to
$\Omega_{tot}$ may only be solved with a paradigm shift, either by
patching of new dark matter+dark energy components or, even more
strikingly, by a deep modification in the description of gravity
fields. Which of these possibilities to choose is difficult to
precise further, because revolutions are complex phenomena and it
is unknown, by definition, which will be the emerging
state-of-the-art.

As a consequence, and with the aim of substantiating this
assertion, we also argue that there is already plenty of evidence
to consider that the 1998 anomaly $\Lambda \neq 0$, taken together
with the "old" dark matter problem has been enough to trigger an
extraordinary science episode as described by Kuhn (K70). Seventy
years of the dark matter problem by itself have not been
uncomfortable enough to do so, and in fact a considerable fraction
of scientists hoped that the dark matter could go away either
because of the identification of some conventional candidate
copiously produced (black holes, brown dwarfs, etc., recently
excluded almost completely using the full set of data of the EROS
experiment$^{23}$) or the detection of a particle candidate that
would have brought dark matter to the realm of everyday physics (a
supersymmetric neutralino, the lightest of the supersymmetry
multiplet, as a prime candidate, see  B. Sadoulet, `Deciphering
the nature of dark matter´. Rev. Mod. Phys., vol. 71, 2000,
S197-S204$^{24}$).

The parallel with the state-of-the-art of physics at the turn of
the 19th century can not be overstated. The astonishing properties
of the ether necessary for a comprehensive understanding of the
classical world did not preclude Lord Kelvin to claim an
essentially complete physical picture in a well-known address to
the British Association for the Advancement of Science$^{25}$.
However, a few years later its complete conceptual elimination and
the paradigm shift to Relativistic and Quantum physics were all
remarkable. Nonetheless, the ether was indeed recognized as a
serious problem by the community and it was attacked fiercely by
several distinguished members (such as Maxwell and Michelson),
thus qualifying as a prototype of the Kuhnian anomaly. If this
parallel is correct, the pair dark matter+dark energy may be truly
considered as the neo-ether of contemporary physics.

Even though not much has been written about how a paradigm shift
actually happens (and there may be several variants), we may
advance here some of the simplest hierarchical possibilities,
namely a top-down or bottom-up path. Typically the "top-down" path
would be the emergence of dark energy, and possibly of dark matter
as well, from a single theory changing quite radically a number of
present sacred concepts. A prototype for the former is brane
theory, which is still "in the works" and which by construction
may harbor new elements contributing to the solution of these
problems$^{26}$. Braneworld models typically embed the 3+1
spacetime in a higher-dimension structure, and as such the
remaining space-like dimensions constitute the "bulk" in which
none of the known elementary forces but gravitation can propagate.
Specific claims about the behavior of braneworld solutions for the
dark matter/dark energy problems have already been made$^{27}$,
the latter fully belonging to the class of extraordinary science
attempts. Conversely, a "bottom-up" path could be taken, starting
for example with phenomenological models (like the Chaplygin gas,
which behaves as dark matter or dark energy in the high and low
density limits$^{28}$) later to be incorporated into a larger
theory but not being merely additive contributions to
Friedmann-Robertson-Walker  cosmology. This would postpone for the
future a physical realization of the phenomenological description
with the identification the elements leading to conceptual breaks.
There is a definite and largely unavoidable possibility that both
approaches, currently being undertaken, can converge in the long
term. Hence, we would recognize after the completion of the
process a new paradigm and its relation to the present one.

To be sure, it was clear how to incorporate $\Lambda \neq 0$ and
dark matter into Friedmann-Robertson-Walker cosmology for years.
But the very existence of dark energy (and dark matter as well) is
what strikes most. The contentious assertions made above apply if
and only if these problems can not be kicked away or brought as
"plug-in" solutions, but rather require an involved reworking of
cosmology.

\section{Features of a paradigm shift: are they being seen?}

Sticking strictly to Kuhn's formulation (K70) of the anomaly
issue, three possible outcomes are foreseen. According to him, the
anomaly is either i) solved by normal science, ii) declared
impossible to solve (because it resisted all radical approaches)
and put aside for a future generation; or iii) triggers the
emergence of a new paradigm and becomes solved within it, becoming
the "normal science" for the next generation.

It is obvious that the combined dark matter+dark energy problem
has not been solved by normal science (this is impressive, even
when considering the very different timescales as recognized
anomalies). It is not clear whether the second alternative can be
actually observed in a finite timescale, in fact, the dark matter
case resisted a few generations of scientists without being "put
aside" at all, at least explicitly. We believe that there are good
reasons for the third alternative to be considered and closely
scrutinized by epistemologists, philosophers of science and
cosmologists/particle physicists alike.

We may also legitimately ask whether the features suggested by
Kuhn as tracers of the state previous to a paradigm shift are also
present in contemporary physics.

$\ast$ First, isolation and characterization of dark matter by
close scrutiny have been achieved, resulting in a pretty good
consensual opinion about the scales in which the latter is present
(galaxies, clusters, etc.); and excluded/allowed regions in the
fiducial mass-cross section plane$^{24}$ and exclusion regions for
astronomical bodies. The efforts to do the same with the much
"newer" dark energy have already resulted in observational limits
intended to pinpoint its exact equation of state and its possible
temporal evolution$^{16}$. The latter also constitute evident
examples of the isolation/characterization processes "in action"
(see Fig. 1), attracting a lot of attention and work. The
excluded/allowed regions and the "equation of state" are clearly
well-defined and acceptable approaches within the idea of dark
matter+dark energy being new components only, as expected from the
existing framework to analyze the data. The features of Fig. 1
serve here to support our view quite directly.

$\ast$ A second feature thought to be indicative of a state
previous to a paradigm shift is the flourishing of
philosophical/methodological analysis. A glimpse at the
specialized literature amply confirms the occurrence of this
feature (to which our very work contributes). This stands in
striking contrast with most disciplines and, more importantly,
with the pre-1998 status, reinforcing our previous statements.

$\ast$ A third signal is thought to be the proliferation of
alternatives, a fact which is also very evident in the literature.
We should also add that the acceleration of this proliferation is
also notorious, although very difficult to track properly. Turner
and Huterer ('Cosmic Acceleration, Dark Energy and Fundamental
Physics', 2007, arXiv:0706.2186$^{29}$) have analyzed some of the
leading solutions today, and it is important to note that all them
have been worked out  after 1998, in attempts to clarify the
situation. Moreover, few people stand for each of these solutions,
as expected for explanations that have yet to prove their
consistency and predictive power.

The remaining two features explicitly discussed by Kuhn as a
prelude to a paradigm shift are of pure psychological nature and
reflect the attitude of the community toward the facts. They are
despair and explicit discomfort. Both are difficult to quantify,
and often expressed only privately (conversations at specialized
meetings, for example). Nevertheless, some explicit examples are
not too difficult to find in the written literature For instance,
the situation has been qualified as "embarrassing" by Rees$^{30}$
and termed "the Kingdom of total ignorance" by de Lima$^{31}$,
among other equally meaningful definitions by leading
cosmologists. These shortcomings are actually in part mitigated by
the visible advance of the knowledge of fundamental parameters ( ,
the value of the Hubble constant, etc., see the relative
contributions of the components of the universe in Fig. 2, which
assumes a "standard" Friedmann-Robertson-Walker  cosmology), and
also by the seizing of a big opportunity to make a relevant
contribution to the field (this being in itself a psychological
factor), but are nonetheless very significant. Overall, we have no
reason to doubt that all the features proposed by Kuhn as
indicating a fertile ground for a paradigm shift are amply
fulfilled nowadays.

\section{Conclusions}

It is not presently known whether the dark matter and dark energy
"problems" are just one or many$^{32}$. The possibility of solving
them by plugging in some alien component into the
Friedmann-Robertson-Walker cosmology + Standard Model of particle
physics is still open, although this solution by itself would
require a modification of the way we think and understand the
content and evolution of the Universe, which would be in itself a
"minor" revolution for cosmology at least, but a major event for
particle physics. There is no firm hint from measured physics
about "dark matter" or "dark energy" particles as yet, and their
existence would open up a whole new physics deeply affecting the
existing view of the microphysical world. The fact that, according
to this possibility, we may be ignoring the composition of $>
95$\% of our universe, and the implication that we are not made of
the same material that most of the universe can not be overstated.

Instead, we may be well inside a true major scientific revolution
in cosmology itself, and thus our vision of the problem still
blurred because precisely of that. This would be the case if full
revision of the way we look at gravitational physics may be needed
(hopefully making dark matter + dark energy go away), as advocated
by some. Particle physics would be pretty much unchanged, but this
outcome would be comparable to the newtonian ? relativistic shift
at the turn of the 20th century.

In both cases an important paradigm shift will be required, and in
fact we have argued here that all the characteristic features of
them, as prescribed by Kuhn, are clearly being seen (wild
proposals, young researchers outside cosmology seizing the
opportunity to contribute, a discomfort inside the cosmologists
community, etc.). We also believe that this "orthodox" behavior
(in the sense of Kuhn) is quite striking, since true scientific
revolutions are complex phenomena for which the original work of
Kuhn description may not be completely adequate. By keeping track
of these and other signals we may be able to witness and
appreciate one of the biggest and rarest events thought to be the
very engine of western science in action.

\section{Acknowledgements}

The author wish to thank the São Paulo State Agency FAPESP for
financial support through grants and fellowships and the partial
supported by CNPq (Brazil). Conversations with G. Lugones, J.A.S.
de Lima, M. Soares-Santos and J.A. de Freitas Pacheco helped to
clarify and refine the views here expressed.

\eject

\section{Figures and Captions}

\bigskip
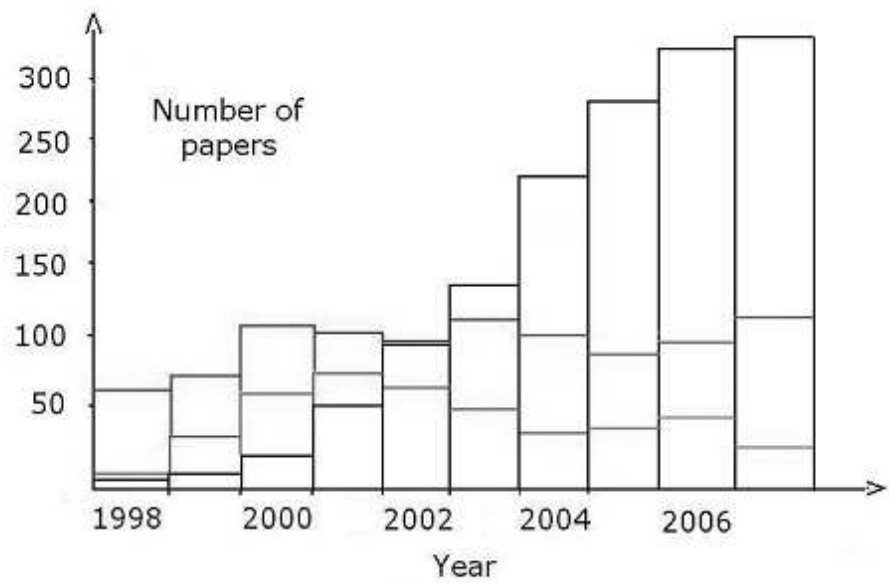
Fig. 1. A graphical representation of the rising interest of the
community in the dark energy problem. This histogram shows the
number of publication having "cosmological constant" (red),
"quintessence" (blue) and "dark energy" (black) in their titles,
collected from the SPIRES/SLAC databases
(http://www.slac.stanford.edu/spires/hep/). While the first two
specific terms remained constant or even declined since 1998, the
more general term "dark energy" grew exponentially, reflecting the
attitude of the community towards the isolation and
characterization of the anomaly. Note that the names of
"quintessence" and "dark energy" did not even exist prior to 1998.

\bigskip
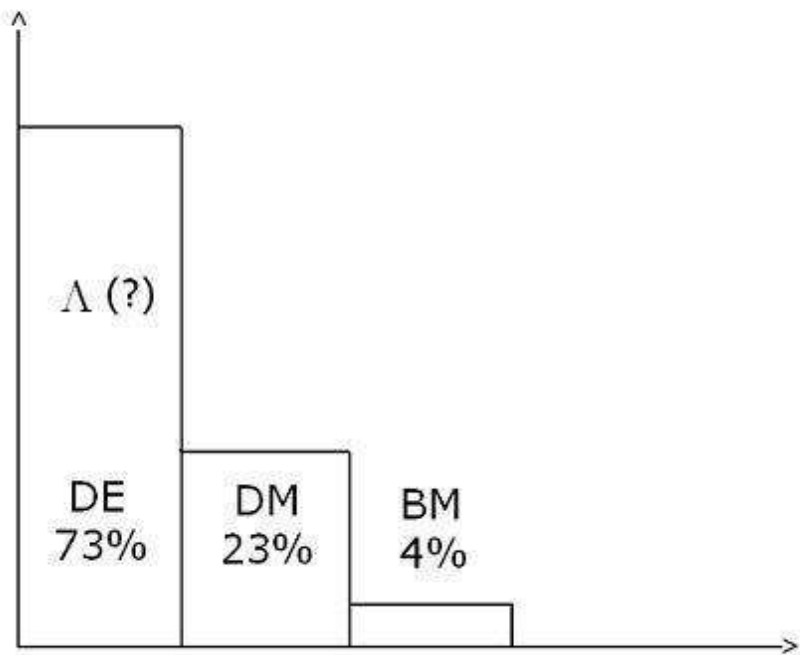
Fig. 2. The most likely content of the Universe according
to the latest observations. The fractions of dark energy, dark
matter and baryonic matter are the best fits to the whole body of
data, and suggests that more of 95 \% of the content of the
universe is unknown.

\end{document}